\documentstyle[12pt,epsf]{article}
\newcommand{\beq}{\begin{equation}}
\newcommand{\eeq}{\end{equation}}
\newcommand{\beqa}{\begin{eqnarray}}
\newcommand{\eeqa}{\end{eqnarray}}
\newcommand{\ba}{\begin{array}}
\newcommand{\ea}{\end{array}}
\newcommand{\CR}{\nonumber \\}
\newcommand{\pa}{\partial}
\newcommand{\A}{\alpha}

\newcommand{\p}{\Phi}
\newcommand{\bra}{\langle}
\newcommand{\ket}{\rangle}

\newcommand{\half}{{1\over 2}}
\newcommand{\Tr}{{\rm Tr}}

\newcommand{\bR}{{\bf R}}

\newcommand{\eq}{\begin{equation}}
\newcommand{\en}{\end{equation}}
\newcommand{\eqn}{\begin{eqnarray}}
\newcommand{\enn}{\end{eqnarray}}

\newcommand{\io}{{\mathbf i }}

\begin{document}

\makeatletter
\def\setcaption#1{\def\@captype{#1}}
\makeatother

\begin{titlepage}
\null
\begin{flushright} 
hep-th/0101087  \\
UT-921 \\
Jan, 2001
\end{flushright}
\vspace{0.5cm} 
\begin{center}
{\LARGE 
A Construction of Commutative D-branes \

from Lower Dimensional Non-BPS D-branes
\par}
\lineskip .75em
\vskip2.5cm
\normalsize

{\large 
Seiji Terashima\footnote{
E-mail:\ \ seiji@hep-th.phys.s.u-tokyo.ac.jp} }
\vskip 1.5em
{\large \it  Department of Physics, Faculty of Science, University of Tokyo\\
Tokyo 113-0033, Japan}
\vskip3cm
{\bf Abstract}
\end{center} \par
We construct an exact soliton, which
represents a BPS D$p$-brane, in a
boundary string field theory action
of infinitely many non-BPS D$(p-1)$-branes.
Furthermore, we show that 
this soliton can be regarded as 
an exact soliton in the full string field theory.
The world-volume theory of 
the BPS D$p$-brane constructed in this way
becomes usual gauge theory on commutative $\bR^{p+1}$,
instead of non-commutative plane.
We also construct a D$p$-brane 
from non-BPS D$(p-n)$-branes by the ABS like configuration.
We confirm that 
the D$p$-brane has correct
RR charges and the tension.

\end{titlepage}

\baselineskip=0.7cm

\section{Introduction}
\setcounter{equation}{0}

The idea of construction of a Dp-brane from 
infinitely many lower dimensional D$(p-2)$-branes has been developed 
in the context of the Matrix theory \cite{BaFiShSu}-\cite{Is}.
Though this is very interesting idea, 
it was known that the Dp-brane which is constructed from 
BPS D$(p-2)$-branes
has charge of the BPS D$(p-2)$-branes
because the lower dimensional brane charges should be conserved.
This means that 
the world-volume of the BPS D$p$-brane 
becomes non-commutative space \cite{CoDoSc}-\cite{Se}.
Therefore it is interesting to construct 
a pure Dp-brane with
commutative world-volume from lower dimensional D-branes.

On the other hand,
some properties of the system containing tachyon
in string theories have been understood \cite{sen18}-\cite{cubic}.
In particular, the non-BPS branes \cite{sen14}-\cite{FrGaLeSt} 
and the brane-antibrane systems \cite{Gr}-\cite{sen13} have been studied.
Furthermore,
the exact solitons
has been found in the effective non-commutative world-volume theories
of unstable D-branes \cite{GoMiSt}-\cite{HaKrLa}.\footnote{
This construction can be viewed as a condensation of the tachyon
which makes the lower dimensional unstable $\infty-1$ 
D$(p-2)$-branes disappear.
A remaining D$(p-2)$-brane corresponds to the 
soliton solution in the non-commutative world-volume theory of 
the unstable Dp-brane
because the infinitely many unstable D$(p-2)$-branes constitute 
the unstable Dp-brane. For more details, see the appendix. }
Here the solitons represent lower dimensional D-branes.
This effective theories can be regarded as the tree level actions
of string field theories,
then, this solitons can be regarded as the exact solutions of 
equations of motion of string field theories.
Recently, even for the case without non-commutativity,
the exact solution has been found 
in the world-volume of the unstable D-branes 
using the boundary string filed theory (BSFT) 
\cite{GeSh1}-\cite{TaTeUe},
which was first formulated by 
Witten \cite{Wi2}. 

Then it is natural to try to construct 
a BPS D-brane from lower dimensional 
unstable non-BPS D-branes which have no conserved charges.
Although some attempts to do this 
was made in \cite{Kl},
the result is not clear, especially for the brane tension, 
since the BSFT action for the non-BPS brane had not been obtained and
the actions used in \cite{Kl}
are not considered to be exact.

In this paper 
we construct an exact soliton, which
represents a BPS D$p$-brane, in the
effective world-volume theory
of infinitely many non-BPS D$(p-1)$-branes.
This world volume theory is obtained from BSFT
and the T-duality. 
In this paper we call a solution of the equations of motion the soliton
although the solution is not localized 
in the world volume of the non-BPS D$(p-1)$-branes.
This is because the solution is localized in the T-dual picture.
Moreover, 
we can show that 
the solution remains intact even
if we include the terms containing many commutators 
in the action.
These terms should be included into the action when we consider
the full string field theory action.
Therefore,
this soliton can be regarded as 
an exact soliton in the string field theory,
as in the construction of the non-commutative soliton \cite{HaKrLa},
where it is argued that derivative corrections,
which correspond to higher commutator terms in our case,
can be ignored.

The world-volume theory of 
the BPS D$p$-brane constructed in this way
becomes usual gauge theory on commutative $\bR^{p+1}$.
The new commutative coordinate of the world-volume of 
the BPS D$p$-brane appears through the 
non-commutative $\bR^2$.
The extra dimension effectively disappears
by the condensation of the tachyons 
as in the case of a soliton in a D$9$-brane,
then the non-commutative $\bR^2$ reduces to the commutative $\bR^1$.

We also construct a D$p$-brane 
from non-BPS D$(p-n)$-branes by the ABS like configuration.
It is shown that the RR charges of the solitons are correct ones.
The tension of the D$p$-brane is shown to
coincide with the expected value for $n=1$ case.
For $n>1$, we show that the tension of the soliton computed from
a natural action coincides with the expected value.

This paper is organized as follows.
In section 2, we briefly review the BSFT action for non-BPS D9-brane
and the construction of the soliton in it.
In section 3,
we construct an exact soliton, which
represents a BPS D$p$-brane, in a
BSFT action
of infinitely many non-BPS D$(p-1)$-branes.
We also construct a D$p$-brane 
from non-BPS D$(p-n)$-branes by the ABS like configuration.
We confirm that 
the D$p$-brane has correct RR charges and the tension.
In section 4, we construct an exact soliton
for brane-antibrane systems.
Section 5 is devoted to conclusion.
In appendix we summarize the construction of 
D$p$-brane from D$(p-2m)$-branes.

\section{Solitons in BSFT action}
\setcounter{equation}{0}

First we review the BSFT action for non-BPS D-brane
and the construction of the soliton in it \cite{KuMaMo2}.
The BSFT action for the real tachyon $T(x)$ 
is obtained in \cite{KuMaMo2} as
\beq
S=\tilde{T}_9 \int d^{10} x e^{-\frac{T^2}{4}} 
F \left( \frac{\A'}{2}\pa_\mu T \pa^\mu T  \right),
\label{bsfta}
\eeq
where 
\beq
F(x)=x \frac{4^x}{2} \frac{\Gamma(x)^2}{\Gamma(2x)}
\label{F}
\eeq
and $\tilde{T}_p$ is the tension of the non-BPS D$p$-brane.
This action is considered to be 
exact if we set $T=\sum u_i x_i$ and other fields to zero.
Using 
\beq
F(x)=1+2 \ln2 \, x +{\cal O}(x^2),
\eeq
we can obtain 
\beq
S=\tilde{T}_9 \int d^{10} x e^{-\frac{T^2}{4}} \left(
1+ \ln2 \, \A'\pa_\mu T \pa^\mu T 
+{\cal O}((\pa_\mu T \pa^\mu T)^2) \right).
\eeq

Then it was shown in \cite{KuMaMo2} that
the closed string vacuum corresponds to $T=\infty$
and the exact soliton which represents a BPS D8-brane
is 
\beq
T=u x_9,
\label{D9sol}
\eeq 
with $u=\infty$
from
\beq
F(x)=\sqrt{\pi x}+\frac{1}{8} \sqrt{ \frac{\pi}{x}}
+{\cal O}(\frac{1}{x^{\frac{3}{2}}}).
\label{fa2}
\eeq
Furthermore, 
the tension of the BPS D8-brane is finite and coincides 
with the expected values.
The fluctuation around  the solution was studied 
in \cite{MiZw} \cite{MiZw2} and 
it was explicitly shown that 
the fluctuation modes live effectively on the ${\bf R}^9$
by the effect of the factor $e^{-T^2/4} \sim e^{-u x_9^2/4}$
which makes the fluctuations along $x_9$ confine in the region
of a length scale $1/\sqrt{u}$.
The spectrum, however, is not expected to be exact because
some neglected terms in the BSFT action, for example 
$e^{-T^2/4} (\pa_\mu \pa_\nu T)( \pa_\mu \pa_\nu T) $, may 
contribute the fluctuation around the solution.

\section{Construction of Dp-branes 
from lower dimensional non-BPS D-(p-2m-1) branes}
\setcounter{equation}{0}

First we consider the soliton 
in a field theoretical toy model for
the effective world-volume action of  
$N$ non-BPS D-branes.
The field theoretical model of 
a tachyon field $T(x)$ and gauge fields $A_\mu(x)$
in a D9-brane considered in \cite{MiZw}
is 
\beq
S=\tilde{T}_9 \int d^{10} x e^{-\frac{T^2}{4}} \left(
1+ c_1 \A' \pa_\mu T \pa^\mu T 
+\frac{c_2}{4} (2 \pi \A' F_{\mu \nu})^2 \right),
\eeq
where $c_i$ are some numerical constants.
Then from the T-dual relation 
\beq
A_{\mu}\rightarrow \frac{1}{\sqrt{\A'}} \p_{\mu},
\eeq
it is natural to consider 
a toy model for $N$ non-BPS D$(-1)$-branes\footnote{In this paper 
we consider the Euclidean 
space-time for simplicity.}
which has an action
\beq
S=\tilde{T}_{-1} \, \Tr_{N \times N} \left[
e^{-\frac{T^2}{4}} I(T,\p)
\right],
\eeq
where $T$ and $\Phi_\mu$ become $N \times N$ matrices
and
\beq
I(T,\p)=
1- c_1 [\Phi_\mu, T]^2 -c_2 
\pi^2  [ \Phi_{\mu} , \Phi_{\nu}]^2 .
\eeq
Here we have done the replacement 
\beqa
D_\mu T=\pa_{\mu} T -i(A_{\mu} T - T A_{\mu}) & \rightarrow &
-i \frac{1}{\sqrt{\A'}} [ \Phi_{\mu} , T] \CR
F_{\mu \nu} =\pa_{\mu} A_{\nu} -\pa_{\nu} A_{\mu} 
-i[A_{\mu},A_{\mu}] & \rightarrow &
-i \frac{1}{\A'} [ \Phi_{\mu} , \Phi_{\nu}].
\eeqa
The equations of motion for this action become
\beqa
0=\frac{\pa S}{\pa T} & \sim &
- I(T,\p)  T e^{-\frac{T^2}{4}}
+4 c_1 
\left[ \p_{\mu} , [\p_{\mu} ,T] e^{-\frac{T^2}{4}} \right] \CR
0=\frac{\pa S}{\pa \p_\mu} & \sim &
-c_1 \left[ T,[\Phi_{\mu} , T] e^{-\frac{T^2}{4}} \right],
\label{emt}
\eeqa
where we have omitted the terms containing 
$[\p_{\mu} ,\p_\nu]$ or $[T,  [\p_{\mu},T ] ]$.

Hereafter we concentrate on the case of 
a system of infinite number of non-BPS D$(-1)$-branes, 
i.e. $N \rightarrow \infty$ limit, and
regard the matrix of $N \times N$ as a linear operator.
Then on the analogy of the solution 
for the case of non-BPS D9-brane (\ref{D9sol}),
we can try to take a configuration
\beqa
T &=& 2 \pi \mu \frac{1}{\sqrt{\A'}} \hat{p} \CR
\p_{0} &=& \frac{1}{2 \pi \sqrt{\A'}} \hat{x}, \hspace{1cm}
\p_{\mu}  =  0  \;\;\; (\mu= 1 \cdots 9),
\label{solt}
\eeqa
where $\hat{x}$ and $\hat{p}$ are operators which satisfy
\beq
[ \hat{x}, \hat{p}] = i \A'.
\eeq
Note that for this configuration, 
it is hold that $[x_{\mu_1},[x_{\mu_1},x_{\mu_1}]]=0$ 
for any choice of assigning $x_{\mu_i}$ to $\p_\nu$ or $T$.
Indeed, this configuration 
becomes a solution for the equations of motion (\ref{emt}) 
provided that $\mu^2= 1/ c_1$ because $[\p_0,T]=i \mu$
and $\frac{\pa S}{\pa T} \sim  -T( 1-c_1 \mu^2)$.
We note that the normalization of the operator $\hat{x}$ 
was taken such that the $\hat{x}$ can be regarded as the 
transverse coordinates of the non-BPS D$(-1)$-branes.
We also note that
the commutator $[ \hat{x}, \hat{p}]$ is taken to depend only on $\A'$.
Actually, a solution for equations of motion 
for general tachyon potentials
was obtained before in \cite{Kl}, however, as a solution of 
differential equation.
We can easily check that 
the solution (\ref{solt}) solves the differential equation.

The tension of this soliton can be computed as 
\beq
S=\tilde{T}_{-1} \Tr \left( e^{-\frac{\pi^2 \mu^2 \hat{p}^2}{ \A'}}
(1+1) \right)=
2 \tilde{T}_{-1} \int dx dp \bra p| e^{-\frac{\pi^2 \mu^2 p^2}{ \A'}}
| x \ket \bra x |p \ket
= \tilde{T}_{-1} \frac{1}{\mu \sqrt{\pi^3 \A'}} \int dx,
\eeq
where we have used the eigenstates $|x \ket$ of operator $\hat{x}$, 
i.e.
$\hat{x} | x \ket=x | x \ket$ with $\bra x | x' \ket=\delta(x-x')$
and the eigenstates of $\hat{p}$ with 
$\bra p | p' \ket=\delta(p-p')$ and $\bra x | p \ket=\frac{1}{\sqrt{2
\pi \A'}} e^{\frac{i p x}{\A'} }$.
Although this soliton is expected to represent a BPS D$0$-brane,
the tension of this soliton does not coincide with
the BPS D0-brane tension 
$T_0=\frac{\tilde{T}_{-1}}{\sqrt{2}} \frac{1}{2 \pi \sqrt{\A'}}$
unless $\mu=\frac{4 }{\sqrt{ 2 \pi}}$.
This is because the action 
can not be regarded as an exact string field theory action.

Now we consider a solution for a BSFT action,
which is derived from the action (\ref{bsfta}) by the T-duality,
\beq
S=\tilde{T}_{-1} \, \Tr_{N \times N} \left[
e^{-\frac{T^2}{4}} \left\{ F \left( -\frac{1}{2} [\p_\mu, T]^2  \right)
+G \left( [\p_\mu,\p_\nu], [\p_\rho, T]   \right)
\right\} \right].
\label{BSFT}
\eeq
This action may be an exact string field theory action for the 
configuration with 
\beq
[T,[\p_\mu, T]]=[\p_\nu,[\p_\mu, T]]=[T,[\p_\mu, \p_\nu]]
=[\p_\nu,[\p_\mu, \p_\rho]]=0,
\label{fr}
\eeq
which correspond to $D_\mu F_{\nu \rho}=0$,
$D_\mu D_\nu T=0$  and $ [T,D_\nu T]=0$. 
Here the function $F$ was defined by (\ref{F}) and
$G(x,y)$ is a some function which is expanded around $x=0$ as 
$G(x,y)=x^2 g(y)+ {\cal O}(x^3)$.
Note that any ordering of $T$, $[\p_\mu,\p_\nu]$ 
and $[\p_\mu, T]$ in (\ref{BSFT}) 
can be taken to have an exact action 
for the configuration which 
satisfies (\ref{fr}).

The equations of motion for this action
become
\beqa
0=\frac{\pa S}{\pa T} & \sim &
-\frac{1}{2} 
F\left( -\frac{1}{2} [\p_\mu, T]^2  \right)  \, T e^{-\frac{T^2}{4}}
-\frac{1}{2} [\p_{\mu} ,T]^2
F'\left( -\frac{1}{2} [\p_\mu, T]^2 \right)
T e^{-\frac{T^2}{4}}    \CR
0=\frac{\pa S}{\pa \p_\mu} & \sim & 0,
\label{emb}
\eeqa
where we have omitted the terms containing 
$[\p_{\mu},\p_\nu]$, $[T,[\p_\mu, T]]$ or $[\p_\nu,[\p_\mu, T]]$.

Then from (\ref{fa2}), 
we can find that 
the configuration (\ref{solt}) is a solution 
for the equation of motion (\ref{emb}) provided that $\mu \rightarrow
\infty$.
Furthermore, the solution (\ref{solt}) is 
an exact solution of the string field theory.
This is because 
the solution (\ref{solt}) satisfies 
$[\p_{\mu},\p_\nu]=0$ and 
$[[\p_\mu, T], H(T,\p_{\nu})]=0$, where $H(x,y)$ is 
an arbitrary function of $x,y$, 
then (\ref{solt}) with $\mu \rightarrow \infty$
remains a solution for 
the equations of motion even if we include
the terms omitted in the action (\ref{BSFT}),
which vanish for the configuration satisfying (\ref{fr}).

The tension of the solution is computed as 
\beq
S=\tilde{T}_{-1} \Tr_{N \times N} 
\left( e^{-\frac{\pi^2 \mu^2 \hat{p}^2}{ \A'}}
F(\mu^2/2) \right)
= \frac{\tilde{T}_{-1}}{\sqrt{2} } \frac{1}{2 \pi \sqrt{\A'}} \int dx,
\eeq
where we have used (\ref{fa2}) and 
\beq
\Tr_{N \times N} \,\left(
e^{-\frac{\pi^2 \mu^2 \hat{p}^2}{ \A'}}  \right)= 
\frac{1}{2 \mu \sqrt{\pi^3 \A'}} \int dx.
\eeq
This indeed coincides with the tension of the 
D$(-1)$-brane.
The Chern-Simons coupling of the non-BPS branes 
which is relevant for this case 
is given by \cite{BiCrRo}-\cite{MuSu} \cite{TaTeUe} as 
\beq
S_{CS} = 
i \mu' \Tr_{N \times N} \left[ \sqrt{\frac{\pi }{2 i}}
[\p_0, T] C_0 e^{-\frac{T^2}{4}} \right],
\eeq
where $\mu'$ is a constant which can be evaluated 
as $\mu'=i^{-\frac{3}{2}} \tilde{T}_{-1}$ by the computation 
of the charge of the 
D${(p-1)}$-brane soliton in non-BPS D$p$-brane BSFT action.
Hence we verify the charge of the soliton as
\beq
S_{CS} =
i^{\frac{3}{2}} \mu' \sqrt{\frac{\pi }{2 }} 
\frac{1}{2 \sqrt{\pi^3 \A'}}  \int C_0 dx,
=\frac{\mu'}{\sqrt{2}} i^{\frac{3}{2}} \frac{1}{2\pi\sqrt{\A'}} 
\int C_0 dx=T_{0} \int C_0,
\eeq
where $T_p$ is a tension of a BPS D$p$-brane.
Note that in this computation we have not taken
the limit $\mu \rightarrow \infty$ 
and the charge does not depend on the value of $\mu$ as expected.

Now we consider the fluctuations around the solution.
First we expect that the fluctuations around the solution
become effectively on the commutative ${\bf R}^1$.
This is because the factor 
\beq
e^{-\frac{T^2}{4}} \sim e^{-\frac{4 \pi^2 }{\A'} \mu^2 \hat{p}^2},
\eeq
with $\mu \rightarrow \infty$ would suppress
fluctuations along the $\hat{p}$.
Note that in the limit $\mu \rightarrow \infty$,
$e^{-\frac{4 \pi^2 }{\A'} \mu^2 p^2}$
reduces to
a delta function $\delta(p)$ in the commutative case
and we can translate the operators to the functions
by the Weyl transformation.
Thus this factor coming from the tachyon condensation
would effectively make the world volume of the 
BPS D0-brane, which is constructed as the soliton,
extend along ${\bf R}^1$, rather than 
non-commutative ${\bf R}^2$.
We note that in this construction of the D0-brane,
the non-commutative ${\bf R}^2$ play a crucial role 
in order to produce the world volume coordinate from the 
matrices.

It is not easy, however, to calculate
the effective action for the fluctuation modes 
since the BSFT actions used in this paper do not 
include the necessary terms to compute these
as mentioned in section 2.
Here we only restrict the fluctuations as
\beqa
T &=& 2 \pi \mu \frac{1}{\sqrt{\A'}} \hat{p} \CR
\p_{0} &=& \frac{1}{2 \pi \sqrt{\A'}} \hat{x}, \hspace{1cm}
\p_{\mu}  =  \delta \p_{\mu}(\hat{x})  \;\;\; (\mu= 1 \cdots 9),
\label{eq0}
\eeqa
i.e. we only consider the $\hat{p}$ independent fluctuations of the 
transverse scalars of the D$0$-brane.
One reason to consider these modes is that 
the modes are decoupled from the other modes up to
quadratic fluctuation.
Indeed, the action should contain the field $\p_{\mu}$ 
as the commutator $[\p_\nu,\p_\mu]$ or $[T,\p_{\mu}]$
and these commutators are already quadratic fluctuation
except
\beq
[T,\p_{\mu}] \sim 2 \pi \mu \frac{1}{\sqrt{\A'}} 
[\hat{p},\p_{\mu}] \sim -2 i \pi \sqrt{\A'} \pa_{\hat{x}} 
(\delta \p_{\mu}( \hat{x})).
\label{eq1}
\eeq
Then inserting (\ref{eq0}) and (\ref{eq1}) into the action (\ref{BSFT}),
we obtain a desired result:
\beq
S = \tilde{T}_{-1} \Tr_{N \times N} \left[ 
F\left(
\frac{\mu^2}{2}
(1+4 \pi^2 \A' (\pa_{\hat{x}} \delta \p_{\mu}( \hat{x}) )^2   )
\right)\right]
=T_0 \int \sqrt{1+4 \pi^2 \A' (\pa_{x} \delta \p_{\mu}( x) )^2 } dx.
\eeq

Of course, we extend the construction of the higher dimensional
D-branes from the infinitely many non-BPS D$(-1)$-branes to
the one from infinitely many non-BPS D$(2m-1)$-branes.
In this case the action for the D$(2m-1)$-branes 
includes the gauge fields. 
Because the analysis is almost same as above discussions, 
we will consider 
more nontrivial generalizations.

We also construct an exact solution, which represents
$n$ BPS D$0$-branes:
\beqa
T &=& 2 \pi \mu \frac{1}{\sqrt{\A'}} 
\left( \begin{array}{c|c|c} \hat{p}  && \\ \hline
& \ddots & \\ \hline
&&\hat{p}  \end{array} \right), \hspace{.1cm}
\p_{0} = \frac{1}{2 \pi \sqrt{\A'}} 
\left( \begin{array}{c|c|c} \hat{x}&& \\ \hline
& \ddots & \\ \hline
&&\hat{x}  \end{array} \right), \hspace{.1cm}
\p_{\mu}  =  
\left( \begin{array}{c|c|c} C_1^{\mu} {\bf 1}  && \\ \hline
& \ddots & \\ \hline
&&  C_{n}^{\mu} {\bf 1}  \end{array} \right), \CR
\label{soltm}
\eeqa
where $\mu= 1 \cdots 9$.
The constants $C_i^{\mu}$ correspond to
positions of the $i$-th brane.
The D0-$\overline{\mbox{D0}}$ brane system 
is obtained by changing signs of $\hat{p}$ in $T$,
for example, $T \sim \left( \begin{array}{c|c} \hat{p}  & 0 \\ \hline
0 & -\hat{p}  \end{array} \right)$.

On the analogy of the ABS configuration \cite{ABS},
a configuration for a D$(m-1)$-brane is expected to be
\beqa
T &=& 2 \pi \frac{1}{\sqrt{\A'}} 
\sum_{k=0}^{m-1} \mu_k \, \gamma_k \, \hat{p}^k, \CR
\p_{i} & = & \frac{1}{2 \pi \sqrt{\A'}} \,
\hat{x}^i \;\;\; (i= 0 \cdots m-1), 
\hspace{1cm}
\p_{\mu}  =  0  \;\;\; (\mu= m \cdots 9),
\label{soltabs}
\eeqa
where $\gamma_i$ are $2^{[m/2]} \times 2^{[m/2]}$ Hermitian 
gamma matrices which satisfy $\{ \gamma_i,\gamma_j \} =2 \delta_{ij}$
and $[ \hat{x}^i, \hat{p}^j] = i \A' \delta_{ij}$.
The D-brane charges of this configuration are expected to vanish 
for $m={\rm even}$, i.e. the non-BPS D$(m-1)$-brane. 
Indeed, we can see that 
the RR charges of the soliton coincide with the expected values.
The Chern-Simons coupling for 
the several non-BPS D$(-1)$-branes was 
obtained by the BSFT \cite{TaTeUe} as,
\beq
S_{CS}=
\tilde{T}_{-1}   i^{-\frac{1}{2}} 
\mbox{SymTr} \left. \left[ e^{-2\pi  i\ \io_{\Phi}\io_{\Phi}+
\sqrt{\frac{\pi  }{2i}}[\io_{\Phi},T]}\wedge C
\, e^{-\frac{1}{4} T^2} \right] \ \right|_{odd},
\label{cs1}
\eeq
where the SymTr denotes a symmetric trace with respect to
$[\Phi_\mu,\Phi_\nu],\ [\Phi_\mu,T]$ and $T^2$
and 
$\io_{\Phi}$ denotes the interior product by ${\Phi}$:
\beq
C=\frac{1}{r!}C_{\nu_1,\nu_2,\cdots,\nu_r} 
dx^{\nu_1}dx^{\nu_2}\cdots dx^{\nu_r}
\rightarrow \,
\io_{\Phi}C=
\frac{1}{(r-1)!} \Phi^i C_{i,\nu_2,\cdots,\nu_r}dx^{\nu_2}\cdots
dx^{\nu_p}.
\eeq
Here $[ \hspace{.5cm} ]|_{odd}$ means that
we should take the odd forms of the RR-fields $C$ only.
Thus we can obtain the RR charges of the soliton as
\beq
C_{CS}=i \frac{1}{\sqrt{2}} \mu' (-i)^{[m/2]} i^{\frac{m}{2}}
\left( \frac{1}{2 \pi \sqrt{\A'}}  \right)^m \int C^{10-m}
=T_{m-1} \int C^{10-m},
\eeq
for $m={\rm odd}$.
Here $\mu'=\tilde{T}_{-1} i^{-\frac{3}{2}} $ and we have used 
${\rm Tr}_{\gamma} (\gamma_1 \cdots \gamma_{2n+1})=2^{n} (-i)^{n}$.

It is difficult, however, to show that
the configuration is a solution
because $[T,[\p_j,T]]=i 2 \pi \frac{1}{\sqrt{\A'}} \mu_j  
\sum_{k=0}^{m-1} \mu_k \, [\gamma_k,\gamma_j] \, \hat{p}^k \neq 0$.
Instead of showing this, 
we consider another action, which seem to be natural,
and compute the tension
using this action.
To this end, let us remember that 
for the configuration 
$T = \sum_{k=0}^{m-1} \mu_k \, \gamma_k \, x^k$,
the BSFT action for $2^{[m/2]}$ non-BPS D9-branes becomes \cite{KuMaMo2}
\beqa
S'  &=& \tilde{T}_{9} \, 2^{[m/2]} \, 
\int d^{10} x\, 
e^{-\frac{T^2}{4}} \prod_{k=0}^{m-1}
\left\{ F \left( \frac{\A'}{2} \mu_k^2  \right)
\right\} \CR
&=& \tilde{T}_{9} \, 2^{[m/2]} \, 
\int d^{10} x\, 
e^{-\frac{T^2}{4}} \prod_{k=0}^{m-1}
\left( 1+ (\ln 2) \A'\mu_k^2 + \beta \A'^2 \mu_k^4+\cdots 
\right)
\CR
&=& \tilde{T}_{9} \, 2^{[m/2]} \, 
\!\!\! \int d^{10} x\, 
e^{-\frac{T^2}{4}} 
\! \left( 1+ (\ln 2) \A' \!\! \sum_{k=0}^{m-1} (\mu_k^2) 
+ \beta \A'^2 \sum_{k=0}^{m-1}(\mu_k^4)
+ \A'^2 \!\! \!\!\!\!
\sum_{k,l=0, \, k > l}^{m-1} \!\!\!\! (\ln 2)^2 \mu_k^2 \mu_l^2
+\!  \cdots \! 
\right)
\CR
&=& \tilde{T}_{9} \, 
\int d^{10} x\, 
e^{-\frac{T^2}{4}} 
\Tr_{\gamma} \left[
1+  (\ln 2) \A' (D_l T)^2  -\half (\ln 2)^2 \A'^2(D_l T D_m T)^2
\right.\CR
&& \hspace{4cm} \left.
+ \left( \beta +
\half (\ln 2)^2 \right) \A'^2 (D_l T)^2 (D_m T)^2 
+\cdots
\right],
\label{BSFT9}
\eeqa
where $\beta$ is a numerical constant
and we have used $\Tr_{\gamma} (\gamma_k \gamma_l \gamma_k \gamma_l )
=2 (\Tr 1) ( 2 \delta_{kl}-1)$.
Then the BSFT action for the $N \times 2^{[m/2]}$ non-BPS D$(-1)$-branes
is expected to be
\beqa
S &=& \tilde{T}_{-1} \Tr_{\gamma} \, \Tr_{N \times N} \left[
e^{-\frac{T^2}{4}} \left(
1-  (\ln 2) [\p_{\mu}, T]^2  
-\half (\ln 2)^2 ([\p_{\mu}, T] [\p_{\nu}, T])^2
\right. \right.\CR
&& \hspace{4cm} \left. \left.
+ \left( \beta +
\half (\ln 2)^2 \right) [\p_{\mu}, T]^2 [\p_{\nu} ,T]^2 
+\cdots
\right)
\right].
\label{acg}
\eeqa
Note that this action may be consistent with the action (\ref{BSFT})
with $N \rightarrow 2^{[m/2]} N$
since (\ref{acg}) may coincides with
(\ref{BSFT}) for $[[\p_\mu,T],[\p_\nu,T]]=0$.
The action (\ref{acg}) is evaluated for the soliton as 
\beq
S=\tilde{T}_{-1} 2^{[m/2]} \, \Tr_{N \times N} \left[
\prod_{k=0}^{m-1} e^{-\frac{\pi^2 \mu_k^2 \, \hat{p}_k^2}{\A'}}
F (\mu_k^2/2)
\right]=T'_{m-1}
\int dx^0 \cdots dx^{m-1},
\eeq
where
\beq
T'_{m-1}=\tilde{T}_{-1} 2^{[m/2]-m} \left(\frac{1}{\sqrt{2 \A'} \pi}
\right)^m.
\eeq
Therefore we have found that 
the $T'_{m-1}$ is coincide with the D$(m-1)$-brane tension
$T^{\rm \, non-BPS}_{m-1}=
(\frac{1}{2 \pi \sqrt{\A'}})^m \tilde{T}_{-1}$ for $m={\rm even}$,
or $T^{\rm BPS}_{m-1}=
(\frac{1}{2 \pi \sqrt{\A'}})^m \frac{\tilde{T}_{-1}}{\sqrt{2}}$ 
for $m={\rm odd}$.

Another exact solution corresponding to BPS D$(2m)$-brane
is an analogue of the usual construction of the 
non-commutative brane as in \cite{Kl}:
\beqa
T &=& 2 \pi \mu \frac{1}{\sqrt{\A'}} \hat{p}, \hspace{1cm}
\p_{0} = \frac{1}{2 \pi \sqrt{\A'}} \hat{x}, 
\CR
\p_{2i-1} \!\!\! &=& \!\! \frac{1}{2 \pi \sqrt{\A'}} \hat{x}_i, 
\;\;\;\;
\p_{2i} = \frac{1}{2 \pi \sqrt{\A'}} \hat{p}_i,  
\hspace{1cm} (i= 1 \cdots m), \CR
\p_{\mu}  &=&  0,   \hspace{1cm}(\mu= 2m+1 \cdots 9),
\label{soltb}
\eeqa
where $\hat{x}_i$ and $\hat{p}_i$ are operators which satisfy
\beq
[ \hat{x}_i, \hat{p}_j] = i \delta_{ij}\A'.
\eeq
As the usual construction of the non-commutative D-branes,
this solution represents the BPS D$(2m)$-brane with
the lower dimensional D-brane charges, 
or background constant NS-NS $B$ field.
This can be verified by the computation of the 
RR charges from the Chern-Simons couplings (\ref{cs1}).
We can also construct the exact soliton 
corresponding to a non-BPS D$(2m-1)$-brane 
with background constant NS-NS $B$ field as 
\beqa
\p_{2i-2} \!\!\! &=& \!\! \frac{1}{2 \pi \sqrt{\A'}} \hat{x}_i, 
\;\;\;\;
\p_{2i-1} = \frac{1}{2 \pi \sqrt{\A'}} \hat{p}_i,  
\hspace{1cm} (i= 1 \cdots m), \CR
\p_{\mu}  &=&  0,   \hspace{1cm}(\mu= 2m \cdots 9),
\;\;\; T=0.
\label{soltb2}
\eeqa
In appendix, we will discuss these cases further.

Finally, we comment on the combinations of the above constructions.
If we notice that the direct sum of the above configurations
is also an exact solution of the equations of motion,
we can easily construct the composite system of 
any number of D$q$-branes and D$(q+m)$ branes.
Therefore it is an advantage of the techniques in this paper that 
we can easily 
construct the system of D-branes with different dimensions.

\section{Brane-antibrane System}
\setcounter{equation}{0}

For D9-$\bar{\rm D}9$ system, 
the BSFT action have been obtained \cite{KrLa} \cite{TaTeUe} as
\beqa
S'  &=& 2 T_{9} \, 
\int d^{10} x\, 
e^{-T \bar{T}}
F \left( \frac{\A'}{2} \mu_1^2  \right)
F \left( \frac{\A'}{2} \mu_2^2  \right)
\CR
&=& 2 T_{9} \, 
\int d^{10} x\, 
e^{-T \bar{T}}
\prod_{k=1}^{2}
\left( 1+ (\ln 2) \A' \mu_k^2 + \beta \A'^2 \mu_k^4+\cdots 
\right)
\CR
&=& 2 T_{9} \, 
\int d^{10} x\, 
e^{-T \bar{T}}
\left( 1+ (\ln 2) \A'  (\mu_1^2+\mu_2^2) 
+ \beta \A'^2 (\mu_1^4+\mu_2^4)
+ \A'^2 (\ln 2)^2 \mu_1^2 \mu_2^2
+  \cdots  
\right)
\CR
&=& 2 T_{9} \, 
\int d^{10} x\, 
e^{-T \bar{T}}
\left[
1+  4 (\ln 2) \A' \pa_l T \pa_l \bar{T}  
+4 \left( 2 \beta-(\ln 2)^2 \right) \A'^2 (\pa_l T \pa_l \bar{T})^2
\right.\CR
&& \hspace{4cm} \left.
+ 4 \left( 2 \beta +
(\ln 2)^2 \right) \A'^2 (\pa_l T \pa_m \bar{T})^2
+\cdots
\right],
\label{BSFT9ddb}
\eeqa
for the configuration 
\beq
T=\half ( i \mu_1 X_1+\mu_2 X_2), \;\;\; A_\mu^{(i)}=0, 
\;\;\; i=1,2.
\eeq
Thus the action for $N$ D(-1)-$\bar{\rm D}(-1)$ pairs is expected to be
\beqa
S \!\! &=& \!\! T_{-1} \Tr_{N \times N} \left[
e^{-T \bar{T}} 
\left\{ 1+ 4 (\ln 2) \A' (\p_l^{(1)} T-T \p_l^{(2)}) 
(\bar{T} \p_l^{(1)} - \p_l^{(2)} \bar{T}) 
\right. \right. \CR
&& \left.\left.  \hspace{.0cm}
+4 \left(  \beta- (\ln 2)^2 \right) \A'^2 (\p_l^{(1)} T-T \p_l^{(2)}) 
(\bar{T} \p_l^{(1)} - \p_l^{(2)} \bar{T}) (\p_m^{(1)} T-T \p_m^{(2)}) 
(\bar{T} \p_m^{(1)} - \p_m^{(2)} \bar{T}) 
\right. \right. \CR
&& \left.\left.  \hspace{.0cm}
+ 4 \left( 2 \beta +
(\ln 2)^2 \right) \A'^2 (\p_l^{(1)} T-T \p_l^{(2)}) 
(\bar{T} \p_m^{(1)} - \p_m^{(2)} \bar{T}) (\p_l^{(1)} T-T \p_l^{(2)}) 
(\bar{T} \p_m^{(1)} - \p_m^{(2)} \bar{T}) 
\right.  \right. \CR
&& \left. \left.  \hspace{.5cm}
\cdots \right\} 
+ e^{- \bar{T} T} 
\left\{ 1+ 4 (\ln 2) \A' (\bar{T} \p_l^{(1)} - \p_l^{(2)} \bar{T}) 
(\p_l^{(1)} T-T \p_l^{(2)}) +\cdots \right\}
\right],
\label{BSFTddb}
\eeqa
where we have omitted $T$-independent terms and higher commutator terms.
For this action we can trace the construction of the higher dimensional
D-branes considered in the section 3.

For example, we can construct an 
exact solution for the infinitely many D(-1)-$\bar{\rm D}(-1)$ pairs
as
\beq
T = \frac{\pi}{\sqrt{\A'}} (i \mu_1 \hat{p}_1 +\mu_2 \hat{p}_2), \;\; 
\p_0^{(1)}=\p_0^{(2)}  =  \frac{1}{2 \pi \sqrt{\A'}} \hat{x}_1 \;\; 
\p_1^{(1)}=\p_1^{(2)}  =  \frac{1}{2 \pi \sqrt{\A'}} \hat{x}_2,
\;\; \mbox{other fields =0},
\eeq
which corresponds to a BPS D1-brane.
A solution which is corresponds to a D$(m-1)$-brane 
can be also constructed for $2^{[m/2]} \times N $
D(-1)-$\bar{\rm D}(-1)$ pairs as 
\beqa
T &=& i \pi \frac{1}{\sqrt{\A'}} 
\sum_{k=0}^{m-1} \mu_k \, \Gamma_k \, \hat{p}^k, \CR
\p_{i}^{(1)}& = & \p_{i}^{(2)}=\frac{1}{2 \pi \sqrt{\A'}} \,
\hat{x}^i \;\;\; (i= 0 \cdots m-1), 
\hspace{1cm}
\;\; \mbox{other fields =0},
\label{solddb}
\eeqa
where $\Gamma^{k}$ denote 
$2^{[\frac{m-1}{2}]} \times 2^{[\frac{m-1}{2}]}$ Hermitian gamma-matrices 
in $m-1$ dimension for $k=1, \cdots, m-1$ and 
$\Gamma^0=i {\bf 1}$.
If we define a shift operator $S$ such as $S^{\dagger} S=1$ and
$S S^{\dagger}=1-P$, where $P$ is a projection operator with 
$\mbox{dim} (\mbox{Ker} P)=N_0$,
a configuration $T=\mu S$ with $\mu \rightarrow \infty$
is an exact solution corresponding to 
$N_0$ D$(-1)$-branes.\footnote{
This solution was suggested by S. Sugimoto.}

\section{Conclusions}
\setcounter{equation}{0}

In this paper 
we have constructed an exact soliton, which
represents a BPS D$p$-brane, in the
effective world-volume theory
of infinitely many non-BPS D$(p-1)$-branes.
We have argued that
the world-volume theory of 
the BPS D$p$-brane constructed in this way
becomes usual gauge theory on commutative $\bR^{p+1}$.
The new commutative coordinate of the world-volume of 
the BPS D$p$-brane is appeared through the 
non-commutative $\bR^2$.
The extra dimension effectively disappears
by the condensation of the tachyons 
as in the case of a soliton in a D$9$-brane.
We have also constructed a D$p$-brane 
from non-BPS D$(p-n)$-branes by the ABS like configuration.
It has been shown that the RR charges and tensions 
of the solitons are correct ones.

It is very interesting to study
the fluctuations around the exact solitons
including the fermion fields to the action as discussed in \cite{MiZw2}.
In order to consider the fluctuations properly, 
we probably have to calculate the 
BSFT action including the higher derivative terms.

\vskip6mm\noindent
{\bf Acknowledgements}

\vskip2mm
I would like to thank K. Ohmori, T. Takayanagi and T. Uesugi
for useful discussions.
I would also like to thank T. Asakawa and S. Sugimoto 
for very stimulating discussions which motivate me to investigate the
subject.
This work was supported in part by JSPS Research Fellowships for Young 
Scientists. \\

\appendix
\setcounter{equation}{0}
\section{D$p$-brane from D$(p-2m)$-branes}

First we consider the construction of 
BPS D$p$-brane from infinitely many BPS D$(p-2)$-branes 
following \cite{BaFiShSu}-\cite{Is}.
For simplicity we consider the construction from BPS D$(-1)$-branes only.
The bosonic part of the effective action for the $N$ BPS D$(-1)$-branes
may be written as 
\beq
S_N=T_{-1} \, {\rm SymTr}_{N \times N} \left[
 \sqrt{ \det \left(  g_{\mu \nu} -i 2 \pi [\p_\mu,\p_\nu]   
\right)}
 \right],
\label{eff1}
\eeq
where we have omitted the terms which contain two or more commutators
and $T_{-1}$ is the tension of a BPS D$(-1)$-brane.
It can be  easily seen that the equations of motion for this action 
with the limit $N \rightarrow \infty$
are satisfied by a configuration 
\beq
\p^{i}=\frac{1}{2 \pi \sqrt{\A'}} \hat{x}^{i}, \,\,
\,\,\; (i= 1,2, \cdots, 2 m ),
\label{s1}
\eeq
where
\beq
[\hat{x}^{j}, \hat{x}^{k}]= i \theta^{jk} \,\,\; (j,k= 1,2, \cdots,2m ),
\eeq
Note that there are moduli $\theta^{jk}$. 
Here we take 
$\theta^{(2j-1), \, 2j} =  -\theta^{2j, \, (2j-1)} \equiv \theta_{j}$
and the other components vanish.

Because we expect that the full effective action contains
$\p_\mu$ only as a commutator $[\p_\mu, \p_\nu]$,
the omitted terms in (\ref{eff1}) can be ignored 
for the configuration (\ref{s1})
as far as the equations of motion is concerned.
Therefore we find that (\ref{s1}) is the exact solution of 
the full string field theory and corresponds to
a D$(2m-1)$-brane with $B_{jk}=(\frac{1}{\theta} )_{jk}$ \cite{SeWi}.
Here following \cite{Se}, we have chosen
$\p_{ij}=-B_{ij}$,
where $\p_{ij}$ represent some degree of freedom 
to write the noncommutative action \cite{SeWi}.

The tension of the solution is evaluated as 
\beqa
S_N &=& T_{-1} \int dx_{1} \int dx_{2} \cdots \int dx_{2m-1} \int dx_{2m}
\sqrt{\prod_{i=1}^m \left( 1+\frac{\theta_i^2}{4 \pi^2 \A'^2} \right)} 
\prod_{j=1}^m \frac{1}{2 \pi \theta_j} \CR
 &=& T_{-1} \left(\frac{1}{2 \pi \sqrt{\A'}}\right)^{2m} 
\prod_{k=1}^{m} \left( 
\sqrt{  1+\left(\frac{2 \pi \A'}{\theta_k} \right)^2}
\int dx_{2k-1} \int dx_{2k}.
\label{ss}
\right)
\eeqa
This coincides with the tension of a D$(2m-1)$-brane since
the world-volume theory of the D$(2m-1)$-brane with non-vanishing 
$B$ field becomes 
\beq
S_{2m-1}=T_{(2m-1)} \left( \prod_{k=1}^{m} 
\int dx_{2k-1} \int dx_{2k}  \right)
\sqrt{ \det \left( g_{ij}+2 \pi \A' B_{ij} \right)},
\eeq
which agrees with (\ref{ss}).

If we denote fluctuations around the solution as
\beq
\p^{i}=\frac{1}{ 2 \pi \sqrt{\A'}} 
\left( \hat{x}^{i} - \hat{A}_{j} \theta^{ji} \right), \,\,
(i=1, \cdots ,2m),
\label{cd}
\eeq
we find \cite{Ao} \cite{Se}
\beq
[\p^i, \p^j]=-i \frac{1}{4 \pi^2 \A'} 
\left(  \theta^{ij} -\theta^{ik} \theta^{lj} \hat{F}_{kl}
\right)= i \frac{1}{4 \pi^2 \A'} \theta^{ik} \theta^{lj} 
\left( \hat{F}_{kl} + \p_{kl}\right),
\eeq
where $\hat{F}_{kl}=\pa_k \hat{A}_l -\pa_l \hat{A}_k
-i [\hat{A}_k, \hat{A}_l]$.
We can also show that 
\beq
(-i)[\p^{i_1},\p_{i_2}] (-i) [\p^{i_2}, \p_{i_3}] \cdots 
(-i)[\p^{i_{2 n-1}},\p_{i_{2n} }] (-i)[\p^{i_{2n}}, \p_{i_1}]=
\Tr ( (G^{-1} (F+\p))^{2n} ),
\eeq
where $G_{ij}= - (2 \pi \A')^2 (B^2)_{ij}$ is the open string metric
\cite{SeWi}.
Using this,
we can show that 
an action for the fluctuations around the solution indeed becomes 
the non-commutative effective world-volume theory
of the D$(2m-1)$-brane
\beq
S_N= T_9 \frac{g_s}{G_s} \sqrt{\det (2 \pi \theta^{ij})} \,
{\rm SymTr}_{N \times N} 
\left( \sqrt{\det(G_{ij}+2 \pi \A' (\hat{F}_{ij}+\p_{ij}}) \right),
\label{eff2}
\eeq
where $G_s=g_s \sqrt{\det(2 \pi \A' B_{ij})}$ 
is the open string coupling constant \cite{SeWi}.
Therefore the action (\ref{eff1}) and (\ref{eff2})
are the same action with the different backgrounds,
which satisfy the equations of motion.

Now we consider the omitted terms in (\ref{eff1}),
which are higher commutator terms.
In principle, we can determine these terms from the non-abelian 
effective world-volume action of D9-branes
by a replacement $2 \pi \A' F_{ij} \rightarrow -i [\p_i, \p_j]$
and $(2 \pi \A')^{\frac{m}{2}+1} 
D_{i_1} \cdots D_{i_m} (F_{kl}) \rightarrow 
-i^{m+1} [\p_{i_1}, [ \cdots ,[\p_{i_m} , [\p_k,\p_l]]]]$.
From (\ref{cd}), we can regard $i \p_i$ by
$\sqrt{2 \pi \A'} \hat{D}_j
= \sqrt{2 \pi \A'} (\hat{\pa}_j+ i \hat{A}_j)$ with the open string metric
in the non-commutative gauge theory picture
where $[\hat{\pa}_j,f(\hat{x},\hat{p})]=\pa_j f(\hat{x},\hat{p})$.
Therefore we find that the full effective action 
for the fluctuations around the solution
has same form as the non-abelian 
effective world-volume action of D9-branes,
as expected.

The D-brane charges can be also evaluated from
the Chern-Simons term \cite{My}
\beq
S_{CS} = T_{-1} {\rm SymTr}_{N \times N} 
\left( e^{ -2 \pi  i \, \io_\p \io_\p } \wedge C \right),
\label{cs}
\eeq
which may be obtained from the Chern-Simons term
for BPS D9-brane 
$S_{CS}^{D9}= T_9 \int d^{10} x e^{2\pi \A' F} \wedge C$.
Substituting the solution into (\ref{cs}),
it is obtained that 
\beq
S_{CS}=  T_{2m-1} \int d^{2m} x \, e^{2\pi \A' B} \wedge C.
\eeq

Now we consider the construction of 
non-BPS D$(2m-1)$-brane from infinitely many non-BPS D$(-1)$-branes.
The action for these was given by (\ref{BSFT}).
The solution for this case was given by (\ref{soltb2})
and the above analysis for the BPS case can be applied 
without essential modifications.
In particular, we can see that 
the noncommutative gauge theory action and 
the matrix action (\ref{BSFT})
are the same action with the different backgrounds,
which satisfy the equations of motion.
In \cite{AgGoMiSt} \cite{HaKrLa}
the non-commutative soliton corresponding to $N_0$ non-BPS D$(-1)$ branes
was constructed in the noncommutative world-volume theory of 
D$(2m-1)$-brane.
It can be identified in the action (\ref{BSFT}) 
as an almost trivial solution
\beq
T=T_c (1-P), \;\; \p_\mu=0, 
\eeq
where $T_c=\infty$ and $P$ is a projector 
which obeys $\mbox{dim} (\mbox{Ker} (1-P))=N_0$.

\newpage


\end{document}